\documentstyle[11pt]{article}
\titlepage
\def\v1{\vspace{1cm}}
\def\be{\begin{equation}}
\def\ee{\end{equation}}
\def\bc{\begin{center}}
\def\ec{\end{center}}
\def\ik{\partial}
\def\vh{\varphi}

\newcommand{\bea}{\begin{eqnarray}}
\newcommand{\eea}{\end{eqnarray}}

\begin{document}
\title
{\bf Proper Time Dynamics in General Relativity and Conformal Unified Theory}

\author{
 L. N. Gyngazov\thanks{Particle
Physics Laboratory, Joint Institute for Nuclear Research,
Dubna,Russia},
M. Pawlowski\thanks{Soltan
Institute for Nuclear Studies, Warsaw, Poland.
; e-mail: pawlowsk@fuw.edu.pl},
V. N. Pervushin
\thanks{Bogolubov Laboratory on Theoretical Physics,
Joint Institute for Nuclear Research, Dubna, Russia},
V. I. Smirichinski
\thanks{Bogolubov Laboratory on Theoretical Physics,
Joint Institute for Nuclear Research, Dubna,
 Russia
; e-mail: smirvi@thsun1.jinr.ru}
}

\date{\empty}

\maketitle
\medskip

\begin{abstract}
{\large
{
The paper is devoted to the description a measurable
time-interval (``proper time'') in the Hamiltonian version of
general relativity with the Dirac-ADM metric.
To separate the dynamical parameter of evolution from the space metric
we use the Lichnerowicz conformally invariant variables.
In terms of these variables GR
is equivalent to the conformally invariant
Penrose-Chernicov-Tagirov theory of a scalar field the role of which is
played by the scale factor multiplied on the Planck constant.

Identification of this scalar field with the modulus of the Higgs field
in the standard model of electroweak and strong interactions allows us
to formulate an example of conformally invariant unified theory
where the vacuum averaging of the scalar field is determined by
cosmological integrals of motion of the Universe evolution. 

}}
\end{abstract}


\newpage

\section {Introduction}

The notion of ``time'', in general relativity, is many-sided
~\cite{MTW,Ryan1,Ryan2}.

General relativity is  invariant 
with respect to general coordinate transformations including the
reparametrizations of the ``initial time-coordinate'' $t\mapsto t'=t'(t)$.

The Einstein observer, in GR, measures the proper time as the
invariant geometrical interval.

The Hamiltonian reduction \cite{MTW} of cosmological models inspired
by GR ~\cite{MTW,Ryan1,Ryan2} reveals
the internal dynamical ``parameter of evolution ''
of the Dirac invariant sector of physical variables~\cite{Dirac,M,KPP,PL1,PL2}.
In cosmological models this ``evolution parameter'' is the cosmic scale
variable, and the
relation between an invariant geometrical interval and dynamical
``evolution parameter'' (the ``proper time'' dynamics) describes
data of the observational cosmology (the red shift and Hubble law).

In this paper we would like to generalize the Hamiltonian reduction 
with internal evolution parameter to the case of field theories of gravity.

For researching the problem of ``time'' in a theory with
the general coordinate transformations
~\cite{MTW}, one conventionally uses ~\cite{Y,K} the
Dirac-ADM parametrization of the metric~\cite{ADM} and
the Lichnerowicz conformally invariant variables ~\cite{AL}
constructed by help of the scale factor
(i.e. the determinant of the space metric).

The Dirac-ADM parametrization is the invariant under
the group of kinemetric transformations.
 The latter contains
 the global subgroup of the reparameterization of time $t\mapsto t'=t'(t)$.
The Hamiltonian reduction of such the time-reparametrization invariant
mechanical systems is
 accompanied by the conversion of one of the initial dynamical variables
 into parameter of evolution of the corresponding reduced systems.
 York and Kuchar \cite{Y,K} pointed out that such variable
in GR
 (which is converted in the evolution parameter) can be proportional
 to the trace of the second form.

 In the contrast with \cite{Y,K}, we suppose that the second form can
 be decomposed on both global excitation and local one.

The ADM-metric and the Lichnerowicz conformally invariant variables
allows us ~\cite{YAF,pap} to extract this
evolution parameter of the reduced system, in GR, as the global component of
the scale factor.

The main difficulty of the Hamiltonian reduction in GR is the necessity of
separation of parameters of general coordinate
transformations from invariant physical variables and 
quantities including the parameter of evolution and proper time.

Recently, this separation was fulfilled in the cosmological Friedmann models
\cite{PL1,PL2}
with the use of the Levi-Civita canonical transformation \cite{LC,PRD,JMP},
which allows one to establish direct relations between the Dirac observables
of the generalized Hamiltonian approach and the Friedmann ones in the
observational cosmology (the red shift and the Hubble law) expressed in terms
of the proper time.

It has been shown that in this way one can construct the normalizable wave
function of the Universe so that the variation of this function under the
proper time leads to the ``red shift'' measured in
observational cosmology \cite{PL2}.

We show that the Hamiltonian reduction of GR distinguishes the conformal time
as more preferable than the proper time from the point of view of the
correspondence principle and causality ~\cite{W}. The usage  of the conformal
time (instead of the proper one) as a measurable interval  can be argued
in the conformal unified theory (CUT) ~\cite{PR,PSP} based on the standard
model of fundamental interactions where the Higgs potential is
changed by the Penrose-Chernicov-Tagirov Lagrangian for a scalar field
~\cite{PCT}.

The content of the paper is the following. In Section 2, we use a model of
classical mechanics with the time reparametrization invariance to introduce
definitions of all times used in the extended and reduced Hamiltonian
systems.  Section 3 is devoted to special relativity to emphasize the main
features of relativistic systems with the frame of reference of an observer.
In Section 4, we consider the Friedmann cosmological models of expanding
Universe to find the relation between the evolution parameter in the
reduced Hamiltonian system and the proper time of the Einstein-Friedmann
observer.  In Section 5, a dynamical parameter of evolution is introduced in
GR as the global component of the space metric, and an equation for the proper
time in terms of this dynamical parameter is derived. Section 6 is devoted to
the construction of conformally invariant theory of fundamental interactions
to analyze similar dynamics of the proper time in this theory.

\section{Classical mechanics}

We consider a reparametrization invariant form of classical mechanics system

\be
\label{2}
W^E\left[p_i,
q_i; p_0,
q_0|t, N\right]=\int\limits_{t_1}^{t_2}dt\left(-p_0\dot q_0+\sum\limits_ip_i
\dot q_i-NH_E(q_0,p_0,q_i,p_i) \right)
\ee
where
\be
\label{2a}
H_E(q_0,p_0,q_i,p_i)=[-p_0+H(p_i, q_i)]
\ee
is the extended Hamiltonian.

The action (\ref{2}) was constructed from 

\be
\label{1}
W^R\left[p_i,
q_i|q_0\right]=\int\limits_{q_0(1)}^{q_0(2)}dq_0\left[\sum\limits_ip_i
\frac{dq_i}{dq_0}-H(p_i, q_i)\right]
\ee
 by the introduction of a
``superfluous'' pair of canonical variables $(p_0, q_0)$ and the Lagrange
factor $(N)$. 

The reduction of the extended system (\ref{2}) to (\ref{1})
means the explicit solution of the equations for ``superfluous'' canonical
variables and the Lagrange factor
\be
\label{3}
\frac{\delta W}{\delta N}=0\,
\Rightarrow\,
-p_0+H(p_i, q_i)=0
\ee
\be
\label{4}
\frac{\delta W}{\delta q_0}=0\,\Rightarrow\, \dot p_0=0
\ee
\be
\label{5}
\frac{\delta W}{\delta p_0}=0\,
\Rightarrow\, dq_0=Ndt\equiv dT.
\ee
Equation (\ref{3}) is a constraint; eq. (\ref{4})
is the conservation law; and eq. (\ref{5}) establishes the relation
between the evolution parameter of the reduced system (\ref{1})
and the ``Lagrange time'', which can be defined for any time 
reparametrization invariant
theory with the use of the Lagrange factor
\be
\label{6}
dT=Ndt.
\ee
The ``Lagrange time'' is invariant $(T(t')=T(t))$.

In the considered case, these two times, $q_0$ and $T$, are equal to each
other due to the equation for ``superfluous'' momenta. However, in the
following, we shall mainly consider opposite cases.

Here, we would like to emphasize that any time reparametrization
invariant theory contains three times: \\
M) the ``mathematical time'' $(t)$ (with a zero conjugate Hamiltonian
(\ref{3}) as a constraint), this time is not observable,\\
L) the ''Lagrange time" $T$ (\ref{6}) constructed with the help of the 
Lagrange factor.\\
D) the dynamical ``parameter of evolution'' of the corresponding
reduced system (\ref{1}), which coincides in this case with the 
``superfluous'' variable
$(q_0)$

The last two times are connected by the equation of motion
for the ``superfluous'' momentum.

\section{Relativistic mechanics}

Let us consider the relativistic mechanics with the extended action
\be
\label{7}
W^E\left[p_i,q_i; p_0,q_0|t, N\right]=
\int\limits_{t_1}^{t_2}dt\left(-p_0\dot q_0+\sum\limits_ip_i
\dot q_i-\frac{N}{2m}[-p^2_0+p_i^2+m^2]\right).
\ee
In this theory, one usually solves the constraint
$-p^2_0+p_i^2+m^2=0$ 
with respect to
the momentum with negative sign in the extended Hamiltonian.
As result we get
\be
\label{10}
\frac{\delta W}{\delta N}=0\,
\Rightarrow\,
(p_0)_{\pm}=\pm\sqrt{p_i^2+m^2},
\ee
so that the conjugate (superfluous) variable converts into the evolution
parameter of the corresponded reduced systems described by
the actions:
\be
\label{pm}
W^R_{(\pm)}\left[p_i,
q_i|q_0\right]=\int\limits_{q_0(1)}^{q_0(2)}dq_0\left[\sum\limits_ip_i
\dot q_i\mp
\sqrt{p_i^2+m^2}\right].
\ee
The latter correspond to two solutions of the constraint.

The variation of action (\ref{7}) with respect to the ``superfluous'' momentum
 $p_0$ gives
\be
\label{8}
\frac{\delta W}{\delta p_0}=-{dq_0\over dt}+N{p_0\over m}=0,
\Rightarrow\,
T_{(q_0)_{\pm}}=\pm\int\limits_0^{q_0}dq_0\frac{m}{\sqrt{p_i^2+m^2}}.
\ee

On the solutions of the equations of motion (\ref{8}) represents 
Lorentz transformation of the proper time $q_0$ of a particle into 
the proper time $T$ of an observer: $T=q_0\sqrt{1-v^2}$.

In this theory we have again three times:\\
M) the ``mathematical time'' $(t)$ (with a zero conjugate Hamiltonian 
as a constraint), this time is not observable,\\
L) the ''Lagrange time" $T$ constructed with the help of the Lagrange
factor and given by (\ref{8}); this time coincides with the proper 
time of an observer.\\
D) the dynamical ``parameter of evolution'' of the corresponding
reduced system (\ref{pm}), which coincides with the proper time of a particle.

In contrast with the mechanical system considered above, the 
evolution parameter (D) differs from the ``Lagrange time" (L) which 
coincides with proper time of the Einstein-Poincar\'e observer. 
The later is defined as the measurable time interval in SR.

\section{ Classical and quantum cosmological models}

We consider the cosmological model inspired by
the Einstein-Hilbert action with an electromagnetic field 
\cite{Ryan1,Ryan2,M,KPP,PL1,PL2}
\be     \label{L1}
W=\int d^4x\sqrt{-g}\left[
-\frac{{}^{(4)}R(g)}{16\pi}M^2_{Pl} -\frac{1}{4}
F_{\mu\nu}(A)F^{\mu\nu}(A)\right]
\ee
If we substitute the Friedmann-Robertson-Walker (FRW) 
metric with an interval
\be \label{L2}
ds^2 = g_{\mu\nu}dx^\mu dx^\nu=a_0^2(t)[N_c^2dt^2-\gamma^c_{ij}dx^idx^j]\;;\;
{}^{(3)}R(\gamma^c)=\frac{6k}{r_0^2}
\ee
into the action, this system  reduces to
the set of oscillators. It is
described by the action in the Hamiltonian form \cite{KPP,PL2}
\be
\label{11}
W^E\left[p_f,
q_f; p_0,
a_0|t, N_c\right]=\int\limits_{t_1}^{t_2}dt\left(-p_0\dot
a_0+\frac{1}{2}\frac{d}{dt}(p_0a_0)
+\sum\limits_fp_f \dot f-N_c\left[-\frac{p^2_0}{4}+{\bf h}^2(a_0)
\right]\right)
\ee
where
\be
{\bf h}^2(a_0)=-\frac{ka^2_0}{r_0^2}+H_M(p_f, f).
\ee
the variable $a_0$ is the scale factor of metric ~(\ref{L2}),
$k=+1, 0, -1$
stands for the closed, flat and open space with the three-dimensional curvature
$(6kr_0^{-2})$. We kept also the time-surface term which follows from the
initial Hilbert action \cite{KPP}.

The equation of motion for the matter ``field'' corresponds to the
conservation law
\be
\label{12}
\frac{d}{dt}H_M(p_f, f)=0
\ee

Let us consider the status of different times (M, L, D) in the theory.\\
M) The main peculiarity of the considered system (\ref{11}) is the invariance
with respect to reparametrizations of the initial time
\be \label{LT}
t\longmapsto  t'=t'(t).
\ee
This invariance leads to the energy constraint and points out that
the initial time $t$ is not observable.\\
L) The ``Lagrange time" $T$ of the extended system (\ref{11}) coincides 
with conformal time $\eta$ \cite{PL2} of the 
Einstein-Friedmann observer who moves together with the Universe 
and measures the proper
time interval $t_F$
\be
\label{13} dt_F=ds|_{dx=0}= a_0N_cdt=a_0d\eta.
\ee
D) The reduction of the extended system (\ref{11}) by resolving the
constraint $\frac{\delta W}{\delta N_c}=0$  with respect to the momentum
with negative sign in the extended Hamiltonian distinguishes the scale
factor as the dynamical parameter of evolution of the reduced system
~\cite{M,Ryan1,Ryan2,KPP}.

The constraint
\be \label{14}
-\frac{p_0^2}{4}+{\bf h}^2=0
\ee
has two solutions
\be
\label{15}
(p_0)_{\pm}=\pm2{\bf h}
\ee
that correspond to two actions of the reduced system
(like in relativistic mechanics considered in Section 3).
The substitution
of (\ref{15}) into eq. (\ref{11}) leads to the action
\be \label{16}
W_{\pm}^R[p_f, f|a_0]=
\int\limits_{a_0(1)}^{a_0(2)}da_0\left[
\sum\limits_f
p_f\frac{df}{da_0}\mp2{\bf h}\pm\frac{d}{da_0}(a_0{\bf h})\right]
\ee
with the evolution parameter $a_0$.

We can see the equation of motion for ``superfluous'' momentum 
$p_0$ of the extended system (\ref{11})
\be
\label{17}
\frac{\delta W}{\delta p_0}=0\,\Rightarrow\,
p_0=2\frac{da_0}{Ndt}=2\frac{da_0}{d\eta}=2a'
\ee
(together with constraint (\ref{15})) establishes the relation between the
conformal and proper times (\ref{13}) of the observer and the
evolution parameter $a_0$ (similar to (\ref{5}) and (\ref{8}))
\be \label{18}
\eta_{\pm}=\pm\int\limits_0^{a_0}da{\bf h}^{-1};~~~~~~~dt_F=a_0(\eta)d\eta.
\ee
Those times can be calculated for concrete values of the integral of motion
\be
\label{19}
H_M=E_c.
\ee
Equation (\ref{18}) presents the Friedmann law \cite{F} of the evolution of
``proper time'' with respect to the ``parameter of evolution''  $a_0$.

{\it The extended system describes the dynamics of the
``proper time'' of an observer with respect to the evolution parameter}.

This proper time dynamics of an
observer of the Universe was used by Friedmann \cite{F}  to describe 
expansion of Universe. This expansion is connected with the Hubble law
\be \label{20}
Z=\frac{a_0(t_F-{D\over c})}{a_0(t_F)}-1\simeq ({D\over c}) {H_{Hub}}(t_F)
+\dots
\ee
where ${H_{Hub}}(t_F)$ is the Hubble parameter and $D$ is the distance 
between Earth and the cosmic object radiating photons.

To reproduce this proper time dynamics the variation principle applied to the
reduced system (\ref{16}) should be added by the convention about measurable
time of an observer (\ref{13}).

In particular,
to get direct relation to the observational cosmology (\ref{20}) of
the Wheeler-DeWitt \cite{WDW} wave function based on the
quantum constraint
\be
\label{21}
\left[-\frac{{\hat p}_0^2}{4}+{\bf
h}^2\right]\Psi_{WDW}(a_0| f)=0\,;\,\left(\hat
p_0=\frac{d}{ida_0}\right)
\ee
equation (\ref{21}) should be added by the convention of an observer
about the measurable time interval (\ref{13}).
In this context,
it has been shown ~\cite{PL2} that there are the Levi-Civita  type canonical
transformations ~\cite{LC} of ``superfluous''
variables $$ (p_0, a_0)\rightarrow(\Pi, \eta) $$ for which the constraint
(\ref{14}) becomes linear
\be \label{22} -\Pi+H_M=0.
\ee
The conformal time of the observer
coincides with the evolution parameter, and the new reduced action
completely coincides with the convential field theory action
of matter fields in the flat space
\be \label{ft}
    W^R_{\pm}[p_f,f|\eta]=
\int\limits_{\eta(1)}^{\eta(2)}d\eta
\left[ \sum\limits_f p_f\frac{df}{d\eta}{\mp} H_M(p_f,f)\right].
\ee
In this case, the WDW equation (\ref{21}) of the new extended 
system coincides with
the Schr\" odinger equation of the reduced system (\ref{ft})
\be \label{23}
\pm\frac{d}{id\eta}\Psi_{\pm}(\eta|f)=H_M\Psi_{\pm}(\eta| f).
\ee
We can get the spectral decomposition of the
wave function of Universe and anti-Universe over ``in'' and ``out'' solutions
and eigenfunctions of the operator $H_M$ with the quantum eigenvalues
$E$ $(H_M<E|f>=E<E|f>)$

\be
\label{24}
\Psi_+(\eta_+|f)=\sum\limits_E\left[
e^{i{\bar
W}^{(+)}_E(\eta_+)}<E|f>\theta(\eta_+)\alpha_{(in)}^{(+)}+
e^{-i{\bar
W}^{(+)}_E(\eta_+)}<E|f>^*\theta(-\eta_+)\alpha_{(out)}^{(-)}\right]
\ee
\be
\label{25}
\Psi_-(\eta_-|f)=\sum\limits_E\left[
e^{i{\bar
W}^{(-)}_E(\eta_-)}<E|f>\theta(\eta_-)\beta_{(out)}^{(-)}+
e^{-i
{\bar
W}^{(-)}_E(\eta_-)
}<E|f>^*\theta(-\eta_-)\beta_{(in)}^{(+)}\right],
\ee
where ${\bar W}^{(\pm)}_E(\eta)$ is the energy part of the reduced actions
(\ref{16}) \cite{KPP,PL2}
\be
\label{26}
W^{(\pm)}_E(\eta_{\pm})=\mp
\int\limits_{a_0(1)}^{a_0(2)}da_0\left[
2{\bf h}-\frac{d}{da_0}(a_0{\bf h})\right]\equiv E\eta_{\pm},
\ee
$\alpha^{(+)}_{(in)}\,,\,\alpha^{(-)}_{(out)}$
 are operators of creation
and annihilation of the Universe $(\Psi_+)$ with the conformal time
$\eta_{(+)}$ and $\beta^{(+)}_{(in)}\,,\,\beta^{(-)}_{(out)}$ are the ones
for the anti-Universe $(\Psi_-)$ with the conformal time $\eta_{(-)}$
(\ref{18}).

If we recall the convention (~\ref{13}) of an observer and
variate the wave function (\ref{24}) with respect to the proper time $t_F$,
we get the red shift energy $E/a_0$ forming the Hubble law. This wave
function has simple interpretation, the same time of evolution as in the
classical theory, and bears direct relation to the observable red shift.

We have got the renormalizable function of the Universe, as we excluded
the superfluous variables from the set of variables of the reduced system.

To obtain this clear quantum theory, we should use the
Einstein-Hilbert action (\ref{L1}), conformally invariant observables, and the
Levi-Civita prescription for the Hamiltonian reduction, which leads to the
conventional matter field theory in the flat space with the
conformal time of an observer.

One can say that the Hamiltonian reduction reveals the
preference of the conformal time from the
point of view of the principle of correspondence with quantum field theory in
the flat space (\ref{ft}) \cite{PL2}.

\section{General relativity}

\subsection{Variables}

The purpose of the present paper is to analyze of  the problem of
``proper time'' dynamics in the exact Einstein-Hilbert-Maxwell theory

\be \label{27}
W^E(g,A)=\int
d^4x\sqrt{-g}\left[-\frac{\mu^2}{6}{}^{(4)}R-\frac{1}{4}F_{\mu\nu}(A)F^{\mu\nu}(A)
\right]\,;~~~\,\left(\mu=M_{Pl}\sqrt{\frac{3}{8\pi}}\right),
\ee
where $M_{Pl}$ is the Planck mass.

The initial points of our analysis are the $(3+1)$ foliation of the
four-dimensional manifold \cite{ADM}
\be
\label{28}
  (ds)^2=g_{\mu\nu}dx^\mu dx^\nu= N^2 dt^2-g^{(3)}_{ij}\breve{dx}{}^i
  \breve{dx}{}^j\;;\;\;(\breve{dx}{}^i=dx^i+N^idt)
\ee
and the Lichnerowicz conformally invariant variables \cite{AL}
\be
\label{29}
N_c=||g^{(3)}||^{-1/6}N;~~~~~g^c_{ij}=||g^{(3)}||^{-1/3}g^{(3)}_{ij};
~~~~~(||g^c||=1);~~~ \bar a=\mu ||g^{(3)}||^{1/6}
\ee
which are convenient for studying the problem of initial data
\cite{MTW,Y,K} and the Hamiltonian dynamics.

With this notation the action (\ref{27}) reads

\be
\label{27a}
W^{E}_{[\bar a, g_c, A]}=\int
d^4x\left[-N_c\frac{\bar a^2}{6}R^{(4)}(g^c)+
\bar a\ik_{\mu}(N_c\ik^{\mu}\bar a)- \frac{1}{4}F_{\mu\nu}(A)F^{\mu\nu}(A)
\right].
\ee

In the first order formalism, the
action (\ref{27}) in terms of the variables (\ref{28}), (\ref{29}) has the
form
\be
\label{30}
W^E=[P_A, A; P_g, g^c, \bar P_a, \bar a|t]=\int\limits_{t_1}^{t_2}dt\int
d^3x\left[\sum\limits_{f=g, A}P_fD_0f-\bar P_aD_0\bar a-N_c
{\cal H}+S\right],
\ee
where
$$
\label{30'}
{\cal H}=-\frac{\bar P_a^2}{4}+6\frac{P_g^2}{\bar a^2}+\frac{\bar a^2}{6}\bar
R+{\cal H}_A
\,;\,\left({\cal H}_A=\frac{1}{2}P_A^2
+\frac{1}{4}F_{ij}F^{ij}\right)
$$
is the Hamiltonian density, $\bar R$ is a three-dimensional curvature
\be
\label{31}
\bar R=R^{(3)}(g^c_{ij})+8\bar a^{-1/2}\Delta\bar a^{1/2}\,;\,\Delta\bar
a=\ik_i(g_c^{ij}\ik_j\bar a),
\ee
$S$ is the surface terms of the Hilbert action (\ref{27}),
$P_A, P_g, \bar P_a $ are the canonical momenta, and
\be
\label{33}
D_0\bar a=\ik_0\bar a-\ik_k(N^k\bar
a)+{2\over 3}\bar a\ik_kN^k\,,\,D_0g^c_{ij}=
\ik_0g^c_{ij}-\nabla_iN_j-\nabla_jN_i+{2\over 3}\ik_k g^c_{ij}N^k 
\ee
\be 
D_0A_i=\ik_0\dot
A_i-\ik_iA_0+F_{ij}N^j
\ee
are the quantities invariant (together with the factor $dt$) under the
kinemetric transformations \cite{YAF}
\be
\label{34}
t\,\rightarrow\, t'=t'(t)\,;\,x^k\,\rightarrow \,x'^k=x'^k(t, x^1, x^2,
x^3) \,,\, N\,\rightarrow\, N'...
\ee
In this theory we also have three ``times''.\\
M)
The invariance of the theory (\ref{30}) under transformations (\ref{34})
(in accordance with our analysis of the problem in the previous Sections)
means that the ``mathematical time" $t$ is not observable.\\
L)
The invariant ``Lagrange time'' defined by the Lagrange factor $N_c$
\be \label{35} dT_c(x, t)=N_c(x, t)dt
\ee
coincides with the measurable proper time in ADM parametrization 
(\ref{28}) within the factor $\bar a/\mu$:
\be
\label{35a}
dT(x,t)=ds|_{dx=0}={\bar a(x,t)dT_c(x,t)\over \mu}.
\ee
D)
The dynamical parameter of evolution of the reduced physical 
sector
as ``superfluous'' variable 
of the extended system (\ref{30}) - a generalization 
of scale factor $a_0$ in cosmology.

For the choice of the ``superfluous'' variable 
in GR we use the results of papers \cite{YAF}
where it has been shown that the space scale $\bar a(x, t)$ contains the global
factor $(a_0(t))$
\be
\label{36}
\bar a(x, t)=a_0(t){\lambda}(x, t)
\ee
which depends only on time and it does not convert into a constant with any
choice of the reference frame in the class of kinemetric 
transformations, where we impose the constraint
\be
\label{37}
\int d^3x{\lambda}(x, t)\frac{D_0{\lambda}(x, t)}{N_c}=0
\ee
which diagonalizes the kinetic term of the action (\ref{30}). 

The new
variables (\ref{36}) require the corresponding momenta $P_0$ and
$P_{\lambda}$.
We define decomposition of $\bar P_a$ over the new momenta $P_0$ and
$P_{\lambda}$
\be
\label{40}
\bar P_a=\frac{P_{\lambda}}{a_0}+
P_0\frac{{\lambda}}{N_c\int d^3x\frac{{\lambda}^2}{N_c}}\,;
~~~~~~~(\int d^3x{\lambda}(x, t)P_{\lambda}\equiv0),
\ee
so that to get the convential canonical structure for the new variables:
\be
\label{38}
\int d^3x(\bar P_aD_0\bar a)=\dot a_0\int d^3x\bar P_a{\lambda}+a_0\int
d^3x\bar P_aD_0{\lambda}
=
\dot a_0P_0+\int
d^3xP_{\lambda}D_0{\lambda}.
\ee
The substitution of (\ref{40}) into the Hamiltonian part of the action
(\ref{30}) extracts the ``superfluous'' momentum term
\be
\label{41}
\int d^3xN_c\bar P_a^2=P_0^2\left[\int
d^3x\frac{{\lambda}^2}{N_c}\right]^{-1}
+\frac{1}{a_0^2}\int d^3xN_cP_{\lambda}^2.
\ee
Finally, the extended action (\ref{30}) acquires the structure of the
extended cosmological model (\ref{11})
\be
\begin{array}{ccr}
W^E[P_f, f;P_0, a_0|t]&=& \int\limits_{t_1}^{t_2}dt\left(\left[\int
d^3x \sum\limits_{f=g_c, A,{\lambda}}P_fD_0f \right] -\dot a_0P_0 \right.
  \\ &&  \\ && \left.+\frac{P_0^2}{4}\left[\int
 d^3x\frac{{\lambda}^2}{N_c}\right]^{-1}-\int d^3xN_c{\cal H}_F\right)
\end{array}
\ee
where ${\cal H}_F$ is
the Hamiltonian ${\cal H}$ without the ``superfluous'' momentum part:
\be \label{43}
{\cal H}_F=\frac{1}{a_0^2}\left[-\frac{P_{\lambda}^2}{4}
+6\frac{P_g^2}{{\lambda}^2}\right]+
a_0^2\frac{\vh^2}{6}\bar R+{\cal H}_A.
\ee
For simplicity we neglect the space-surface term.

\subsection{Reduction}

Now we can eliminate the ``superfluous'' variables $a_0, P_0$
resolving the constraint
\be
\label{45}
\int d^3xN_c\frac{\delta W}{\delta N_c}=0\,\Rightarrow\,
\frac{P_0^2}{4}=\left(\int
d^3xN_c{\cal H}_F\right)\left(\int d^3x\frac{{\lambda}^2}{N_c}\right).
\ee
with respect to the momentum $P_0$.
This equation has two solutions that correspond to two reduced systems with
the actions
\be
\label{46}
W^R_{\pm}(P_f,
f|a_0)=\int\limits_{a_0(1)}^{a_0(2)}da_0\left[\sum\limits_{f={\lambda},g_c,A}
P_fD_af\mp\left(\int d^3xN_c{\cal H}_F\right)^{1/2}\left(\int
d^3x\frac{{\lambda}^2}{N_c}\right)^{1/2})\right]
\ee
with the parameter of evolution $a_0$, where
\be
\label{47}
D_af=\frac{D_0f}{\dot a_0}
\ee
is the covariant derivative with the new shift vector $N^k$ and vector
field $A_{\mu}$, which differs from the old ones, in (\ref{33}), by the factor
$(\dot a_0)^{-1}$.

The local equations of motion of systems (\ref{46}) reproduce
the invariant sector of the initial extended system and determine
the evolution of all variables $(P_f, f)$ with respect to the parameter
$a_0$
\be
\label{48}
(P_f(x, t), f(x, t),\dots)\,\rightarrow\,(P_f(x, a_0), f(x, a_0),\dots).
\ee
The actions (\ref{46}) are invariant under the transformations $N_c(x,
t)\,\rightarrow\,N'_c=f(t)N_c$.
In other words, the lapse function $N_c(x, t)$ can be determined up to
the global factor depending on time:
\be
\label{49}
N_c(x, t)=N_0(t){\cal N}(x, t)
\ee
This means that the reduced system looses
the global part of the lapse function which forms the global time of an
observer
\be \label{50}
N_0dt=d\eta\,;~~~~~~~~~~\,(\eta(t')=\eta(t))
\ee
like the reduced action of the cosmological model lost the lapse function
which forms the conformal time of the Friedmann observer of the evolution of
the Universe considered in the previous Section).

We call quantity  (\ref{50}) the global conformal time.  We can
define the global lapse function $N_0(t)$ using the second integral in eq.
(\ref{45})
\be \label{51} \int d^3x\frac{{\lambda}^2}{N_c}\buildrel{\rm
def}\over=\frac{l_0}{N_0(t)}
\ee
where $l_0$ is the constant which can be
chosen so that ${\cal N}(x, t)$ and ${\lambda}(x,t)$
in the Newton approximation have the form
\be
\label{52}
{\cal N}(x, t)=1+\delta_N(x)+\dots;~~~~~~~
{\lambda}(x, t)=\mu (1+\delta_{\lambda}(x)+\dots
\ee
where $\delta_N$, $\delta_{\lambda}(x)$ are the potentials
of the Newton gravity.

\subsection{The proper time dynamics}

To research the evolution of the system with respect to the global
conformal time of an observer (\ref{50}), we shall use the short notation
\be
\label{53}
\int d^3x N_c{\cal H}_F=l_0N_0{\bf h}^2(a_0)\equiv l_0N_0\left[
\frac{{\bf
k}_A^2}{a_0^2}+{\bf h}_R^2+a_0^2{\bf \Gamma}^{-2}
\right]
\ee
where ${\bf k}_A^2$ and ${\bf \Gamma}^{-2}$ correspond to the kinetic and
potential parts of the graviton Hamiltonian in eq. (\ref{43}), ${\bf h}_R^2$
is the electromagnetic Hamiltonian.

The equations for superfluous variables $P_0$, $a_0$ and global 
lapse function (which are omitted by the reduced
action (\ref{46})) have the form
\be
\label{54}
N_0
\frac{\delta W^E}{\delta N_c}=0\,\Rightarrow\,
(P_0)_{\pm}=\pm2l_0{\bf h}(a_0)
\ee
\be
\label{55}
\frac{\delta W^E}{\delta a_0}=0\,\Rightarrow\, P'_0=l_0\frac{d}{da_0}{\bf
h}^2(a_0);~~~~~~~~(f'=\frac{d}{d\eta}f)
\ee
\be
\label{56}
\frac{\delta W^E}{\delta P_0}=0\,\Rightarrow\, a'_0=\frac{P_0}{2l_0}
\ee
These equations lead to the conservation law
\be
\label{57}
\frac{({\bf k}_A^2)'}{a_0^2}+({\bf h}_R^2)'+a_0^2({\bf \Gamma}^{-2})'=0
\ee
and to the Friedmann-like evolution of global conformal time of an
observer (\ref{50})
\be \label{58}
\eta_{(\pm)}(a_0)=\pm\int\limits_0^{a_0}da{\bf h}^{-1}(a).
\ee
The integral (\ref{58}) can be computed, if we know a
solution of the reduced system of equations (\ref{48}) as functions of the
parameter of evolution $a_0$. To get this equations, we should change, in eq.
(\ref{46}), $N_c$ by ${\cal N}(x, t)$ (as we discussed above).

The conservation law (\ref{57}) allows us to verify that the
red shift and the Hubble law for our observer
\be
\label{59}
Z(D)=\frac{a(t_F)}{a(t_F-D)}-1=D\cdot H_0+\dots ;~~~~~~~~\left(t_F(\eta)=
\int\limits_0^{\eta}d\eta'a_0(\eta')\right)
\ee
reproduce the evolution of the Universe in the standard cosmological models
(with the FRW metrics), if we suppose the dominance of the kinetic part of
the Hamiltonian or the potential one, in accordance with the $a_0$-dependence
of this Hamiltonian.

In the first case
$({\bf k}_A^2\ne0, {\bf h}_R={\bf\Gamma}^{-1}=0)$,
 we get
the Misner anisotropic model \cite{M} in the second case, the Universe
filled with radiation $({\bf k}_A^2=0; {\bf h}_R\ne0; {\bf
\Gamma}^{-1}\ne0)$. In both the cases, the quantities
$({\bf k}_A, {\bf h}_R,{\bf\Gamma}^{-1})$ play the role of conserved
integrals of motion which are constants on solutions of the
local equations.

The ``Lagrange time" differential (\ref{35}) is 
\be
\label{35b}
dT_c(x,t)={\cal N}(x,\eta)d\eta.
\ee

In the quantum theory, the integrals of motion become conserved quantum
numbers (in accordance with the correspondence principle). Each term of the
spectral decomposition of the wave function over quantum numbers can be
expressed in terms of the proper time of an observer to distinguish ``in''
and ``out'' states of the Universe and anti-Universe with the corresponding
Hubble laws.

Attempts \cite{PL2} to include an observer into the reduced scheme (by
the Levi-Civita canonical transformation \cite{LC,PRD,JMP}
of the extended system
variables to the new ones for which the new ``superfluous'' variable
coincides with the proper time) show that the conformal time and space
observables are more preferable than proper time and space. The conformal
time leads, in the flat space limit, to the quantum field theory action
\cite{KPP} and does not violate causality \cite{W} (in contrast with
the proper one). The conformal space interval does not
contain singularity at the beginning of time \cite{KPP,PL1,PL2}.  In the next
Section we try to remove these defects changing only the convention of
measurable intervals and keeping the physics of the reduced system unviolated.

\section{Conformal Unified Theory (CUT)}

\subsection{The formulation of the theory}

Our observer in his (3+1) parametrization of metric
can see that the Einstein-Hilbert
theory,
in terms of the Lichnerowicz conformally invariant variables (\ref{29}),
completely coincides with the conformal invariant theory of the
Penrose-Chernicov-Tagirov (PCT) scalar field with the action (except the
sign)
\be \label{60} W^{PCT}{[\Phi, g]}=\int
d^4x\left[-\sqrt{-g}\frac{\Phi^2}{6}R^{(4)}(g)+
\Phi\ik_{\mu}(\sqrt{-g}\ik^{\mu}\Phi)\right],
\ee
if we express this action also
in terms of the Lichnerowicz conformally invariant variables:
\be
\label{61}
\vh_c=
||g^{(3)}||^{1/6}
\Phi\,;\,g^c_{\mu\nu}=||g^{(3)}||^{-1/3}g_{\mu\nu}\,;\,\sqrt{-g^c}=N_c.
\ee
From (\ref{60}) we get the action
\be
\label{62}
W^{PCT}{[\vh_c, g_c]}=\int
d^4x\left[-N_c\frac{\vh_c^2}{6}R^{(4)}(g^c)+
\vh_c\ik_{\mu}(N_c\ik^{\mu}\vh_c)\right].
\ee
which coincides with the Einstein action in eq. (\ref{27a}) if we replace 
$\bar a$ with $\vh_c$.
However,
in contrast with the Einstein theory, the observables in PCT theory
are conformally invariant quantities, in particular, an observer measures
the conformally invariant interval
\be
\label{64}
  (ds)_c^2=g^c_{\mu\nu}dx^\mu dx^\nu= N^2_cdt^2-g^{(3)c}_{ij}\breve{dx}{}^i
  \breve{dx}{}^j
\ee
with the conformal time $\eta$ and  the conserved volume of the
conformally invariant space (as $||g^{(3)c}_{ij}||=1$).

Following refs. \cite{BM,PR,PSP,2}, we
can identify the PCT-scalar field with the modulus of the Higgs
dublet and add the matter fields as the conformally invariant part of the
standard model (SM) for strong and electroweak interactions with the action
\be
\label{65}
W^{SM}[\phi_{Hc},n,V,\psi,g_c] =\int d^4x\left({\cal
L}_0^{SM}+N_c[-\vh_{Hc}F+\vh_{Hc}^2B-\lambda\vh_{Hc}^4]\right),
\ee
where ${\cal L}_0^{SM}$ is the scalar field free part of SM expressed in
terms of the conformally invariant variables of the type of (\ref{61})
\cite{PSP}, $B$ and $F$ are the mass terms of the boson and fermion fields
respectively:
\be
\label{66}
B=Dn(Dn)^*\,;\,F=(\bar\psi_Ln)\psi_R+ h. c.
\ee
They can be expressed in terms of the physical fields $(V_i^p,
\psi_{\alpha}^p)$, in the unitary gauge,
\be \label{67}
B=V_i^p\hat Y_{ij}V_j^p\,;\,F=\bar\psi_{\alpha}^p\hat X_{\alpha\beta}\psi_{\beta}^p
\ee
which absorb the angular components $(n)$ of scalar fields
(here $\hat Y_{ij}, \hat X_{\alpha\beta}$ are the matrices of coupling
constants).

We have introduced the rescaled scalar field $\vh_{Hc}$
\be
\label{67a}
\vh_{Hc}=\chi\vh_c
\ee
in order to ensure a correspondence with ordinary SM notation. 
The rescaling factor $\chi$ must be regarded as a new coupling constant 
which coordinates weak and gravitational scales \cite{PR}. (The value of $\chi$ 
is very small number of order of ${m_W\over M_{Pl}}$ where $m_W$ is the mass of 
weak boson $W$.)

The conformally invariant unified theory (CUT) of all fundamental interactions
\be
\label{68}
W^{CUT}[\phi_c,V^p,\psi^p,g_c] =W^{PCT}[\phi_c,g_c] +W^{SM}[\phi_c,V^p,\psi^p,g_c]
\ee
does not contain, in the Lagrangian, any dimensional parameters.

\subsection{Reduction}

We can apply, to CUT, the analysis of the notions of
 ``times '' in the previous Sections.

The scalar field in CUT acquires the feature of the scale factor component of
metric with the negative kinetic energy and the evolution parameter
$a_0$ can be extracted from the scalar field.
It is convenient to use for global component the
denotations
\be
\label{d1}
\varphi_c(x, t)=\varphi_0(t)a(x, t);~~~~~~N=N_0(t){\cal N}(x, t)
\ee
so that the expression for the extended action has the form
\be
\label{d3}
W^{CUT}(P_f, f; P_0, \varphi_0| t)=\int\limits_{t_1}^{t_2}\left(\int
d^3x\sum\limits_{f=a, g_c,
F_{SM}}P_fD_0f-P_0\dot\varphi_0-N_0\left[-\frac{P_0^2}{4V_0}
+H_f[\varphi_0]\right]\right)dt,
\ee
where $F_{SM}$ is the set of the SM fields,
\be
\label{d4}
H_f[\varphi_0]=
\int d^3x{\cal N}{\cal H}(p_f, f, \varphi_0)={\bf h}_{CUT}^2(\varphi_0)V_0,~~~~~~~
V_0=\int d^3x\frac{a^2}{{\cal N}}
\ee
is the
Hamiltonian of the local degrees of freedom, the Newton perturbation theory
for $a, {\cal N}$ begins from unit $(a=1+\dots\,,\,{\cal N}=1+\dots)$, (the
time-surface term is omitted).

The reduction means that we consider the extended action (\ref{d3}) onto the
constraint
\be
\label{d6}
\frac{\delta W^E}{\delta N_0}=0.\,\Rightarrow\,
(P_0)_{\pm}=\pm2\sqrt{V_0 H_f}.
\ee
The reduced action
\be
\label{d7}
W^R_{\pm}(P_f, f|
\varphi_0)=\int\limits_{\varphi_1=\varphi_0(t_1)}^{\varphi_2=\varphi_0(t_2)}
d\varphi_0 \left\{\left(\int
d^3x\sum\limits_fP_fD_{\varphi}f\right)\mp2\sqrt{V_0 H_f}\right\}
\ee
is completed by the proper time dynamics.

\subsection{The proper time dynamics}

The equations of global dynamics
(which are omitted by the reduced
action (\ref{d7})) have the form
\be
\label{d54}
\frac{\delta W^E}{\delta N_0}=0\,\Rightarrow\,
(P_0)_{\pm}=\pm2V_0{\bf h}_{CUT}({\varphi_0})
\ee
\be
\label{d55}
\frac{\delta W^E}{\delta {\varphi_0}}=0
\,\Rightarrow\, P'_0=V_0\frac{d}{d{\varphi_0}}{\bf h}^2_{CUT}({\varphi_0})
;~~~~~~~~(f'=\frac{d}{d\eta}f)
\ee
\be
\label{d56}
\frac{\delta W^E}{\delta P_0}=0\,\Rightarrow\,
\left(\frac{d\varphi_0}{d\eta}\right)_{\pm}
=\frac{(P_0)_{\pm}}{2V_0}
=\pm{\bf h}_{CUT}({\varphi_0}),
\ee
where the effective Hamiltonian density functional has the form
\be
\label{71}
{\bf h}^2_{CUT}=
\frac{{\bf k}_A^2}{{\varphi_0}^2}+{\bf
h}_R^2+{\bf\mu}_F^2{\varphi_0}
+{\bf\Gamma}_B^{-2}{\varphi_0}^2+{\bf\Lambda}{\varphi_0}^4,
\ee
in correspondence with the new terms in the CUT action.

These equations lead
to the Friedmann-like evolution of global conformal time of an
observer
\be
\label{70}
\eta({\varphi_0})=\int\limits_0^{\varphi_0}d\varphi{\bf h}^{-1}_{CUT}(\varphi),
\ee
and to the conservation law
\be
\label{72}
\frac{({\bf k}_A^2)'}{{\varphi_0}^2}+
({\bf h}_R^2)'+({\bf\mu}_F^2)'{\varphi_0}+
({\bf\Gamma}_B^{-2})'{\varphi_0}^2+({\bf\Lambda})'{\varphi_0}^4=0.
\ee
The red shift and the Hubble law in the conformal time version
\be
\label{73}
z(D_c)=\frac{{\varphi_0}(\eta_0)}{{\varphi_0}(\eta_0-D_c)}-1\simeq
D_cH_{Hub};~~~~~~~~~~\,H_{Hub}
=\frac{1}{{\varphi_0}(\eta)}\frac{d}{d\eta}{\varphi_0}(\eta)
\ee
reflects the alteration of size of atoms in the process of evolution of
masses ~\cite{Narlik,PL2}.

In the dependence on the value of ${\varphi_0}$, there is dominance of the
kinetic or the potential part of the Hamiltonian (\ref{71}), (\ref{72}) and
different stages of evolution of the Universe (\ref{70}) can appear:
anisotropic $({\bf k}_A^2\ne0)$ and radiation $({\bf h}_R^2\ne0)$ (at the
beginning of the Universe), dust $({\bf\mu}_F^2\ne\,;\,{\bf\Gamma}_B^{-2})$
and De-Sitter $\Lambda\ne0$ (at the present time).

In perturbation theory, the factor $a(x,t)=(1+\delta_{a})$
represents the potential of the Newton gravity $(\delta_{a})$.
Therefore, the Higgs-PCT
field, in this model, has no particle-like excitations
(as it was predicted in paper \cite{PR}). 

\subsection{Cosmic Higgs vacuum}

Let us show that
value of the scalar field in CUT
is determined by the present  state of
the Universe with observational density of matter
${\rho_{Un}}$ and the Hubble parameter $H_{Hub}$.

For an observer, who is living in the Universe,
a state of ``vacuum'' is the state of the Universe at present time:
$|Universe>=|Lab.vacuum>$,
as his unified theory pretends to describe both observational cosmology and
any laboratory experiments.

In correspondence with this definition, the Hamiltomian (\ref{d4})
can be  split into the large (cosmological -- global) and small (laboratory
-- local) parts
\be\label{d5}
H_f[\varphi_0]
\buildrel{\rm
def}\over
=
\rho_{Un}V_0+(H_f-\rho_{Un}V_0)=
\rho_{Un}(\varphi_0)V_0+H_L
\ee
where the global  part of the
Hamiltonian $\rho_{Un}(\varphi_0)V_0$ can be defined as the ``Universe''
averaging
\be\label{d27}
<Universe|H_f|Universe>=\rho_{Un}V_0,
\ee
so that the ``Universe'' averaging of the local part of Hamiltonian (\ref{d5})
is equal to zero
\be\label{d28}
<Universe|H_L|Universe>=0.
\ee

Let us suppose that the local dynamics $(H_L)$ can be neglected if we
consider the cosmological sector of the proper time dynamics (\ref{d54}),
(\ref{d55}), (\ref{d56})
\be
\label{d9}
\frac{\delta W^E}{\delta N_0}=0.\,\Rightarrow\,
p_0=2V_0
\sqrt{\rho_{Un}+\frac{H_L}{V_0}}=
2V_0\sqrt{\rho_{Un}}+\frac{H_L}{\sqrt{\rho_{Un}}}+
o\left(\frac{1}{V_0}\right)
\ee
\be
\label{d10}
\frac{\delta W^E}{\delta P_0}=0\,\Rightarrow\,
\left(\frac{d\varphi_0}{d\eta}\right)_{+}=\sqrt{\frac{H_f}{V_0}}
=
\sqrt{\rho_{Un}+\frac{H_L}{V_0}}=\sqrt{\rho_{Un}}
+o\left(\frac{1}{V_0}\right).
\ee
The evolution of the proper time of an observer with
respect to the evolution parameter $\varphi_0$ determines
 the Hubble ``constant''
\be
\label{d11}
H_{Hub}=\frac{1}{\bar\varphi_0(\eta_0)}\frac{d\bar\varphi(\eta_0)}{d\eta_0}=
\frac{\sqrt{\rho_{Un}(\varphi_0)}}{\bar\varphi_0(\eta_0)}.
\ee
The last equality follows from eq.(\ref{d10}) and gives the relation
between the present-day value of scalar field and
the cosmological observations:
\be
\label{d12a}
\bar\varphi(\eta=\eta_0)=\frac{\sqrt{\rho_{Un}(\eta_0)}}{H_{Hub}(\eta_0)}.
\ee

If
$\rho_{Un}=\rho_{cr}$, where
\be
\label{d13}
\rho_{cr}=\frac{3H_{Hub}^2M_{Pl}^2}{8\pi},
\ee
as it is expected in the observational cosmology, 
then the substitution of (\ref{d13}) into (\ref{d12a})
leads to the value of scalar field
\be
\label{d12b}
\bar\varphi(\eta=\eta_0)=M_{Pl}\sqrt{\frac{3}{8\pi}},
\ee
what corresponds to the Newton coupling constant in Einsten's theory of gravity.

%
%
%

\subsection{The dust Universe}

The present-day Universe is filled in by matter
with the equation of state of the dust at rest. This means
the ``vacuum''
averaging of the mass term in the SM Hamiltonian is equal to the mass of the
Universe $M_D$, while other terms can be neglected:
\be\label{d20}
\rho_{Un}V_0=
\varphi_0(\eta)<Univ.|
\int\limits_Vd^3x{\cal N}a
\bar\psi_{\alpha}X_{{\alpha}{\beta}}\psi_{\beta}|Univ.>
\buildrel{\rm def}\over\equiv M_D
=\varphi_0(\eta)<n_b>V_0,
\ee
where
$<n_b>$ is the conserved integral of motion.
In this case, the proper time dynamics is described by eq.
(\ref{d10}) with the density
\be\label{d23}
\rho_{Un}(\varphi_0)=\varphi_0<n_b>;~~~~~~~~
\frac{d\varphi_0}{d\eta}=\sqrt{\varphi_0 <n_b>}.
\ee
We get the evolution law for a scalar field
\be\label{d24}
\varphi_0(\eta)=\frac{\eta^2}{4}<n_b>
\ee
and the Hubble parameter $H_{Hub}(\eta)$
\be\label{d25}
H_{Hub}=\frac{1}{\varphi_0}\frac{d\varphi_0}{d\eta}=\frac{2}{\eta} .
\ee
The barion density
\be\label{d30}
\rho_b=\Omega_0\rho_{Un};~~~~~~~~~~~~~(\rho_{Un}=
\frac{3H_{Hub}^2M_{Pl}^2}{8\pi})
\ee
is estimated from experimental data on
luminous matter ($\Omega_0=0.01$), the flat rotation curves of spiral
galaxies ($\Omega_0=0.1$) and others data~\cite{rpp} ($0.1<\Omega_0<2$).

We should also take into account that these observations
reflects the density at the time of radiation of
 a light from cosmic objects $\Omega(\eta_0-distance/c)$
which was less than at the present-day density $\Omega(\eta_0)=\Omega_0$
due to increasing mass of matter.
This effect of retardation can be roughly estimated by the averaging
of $\Omega(\eta_0-distance/c)$ over distances (or proper time)
\be\label{d31}
{\gamma}=\frac{\eta_0 \Omega_0}{\int\limits_0^{\eta_0} d\eta\Omega(\eta)}.
\ee
For the dust stage the coefficient of the increase is $\gamma=3$.
Finally, we get the relation of the cosmic
value of the Planck ``constant'' and the GR one.
\be\label{d32}
\frac{\bar\varphi(\eta=\eta_0)}{M_{Pl}}\sqrt{{8\pi\over 3}} 
=\sqrt{\gamma\Omega_{0(exp)}}/h=\omega_0,
\ee
where $h=0.4\div 1$ is observational bounds for the Hubble parameter.

From data on $\Omega_0$ we can estimate $\omega_0$: $\omega_0=0.04$ 
(luminous matter), $\omega_0=0.4$ (flat rotation curves of spiral
galaxies), and $0.4<\omega_0<9$ (others data~\cite{rpp}) for lower values of
$h$ ($h=0.4$).

\subsection{The local field theory}

As we have seen in cosmological models there is a Levi-Civita 
canonical transformation to new variables for which the 
``Lagrange time" coincides with the evolution parameter and the 
extended system converts into a conventional field theory.
In general case it is difficult to find the exact form of this LC 
transformation. However, we can proof the equivalence of our reduced 
system with conventional field theory with measurable conformal 
time in next order of the expansion in $V_0^{-1}$ (i.e. the inverse 
volume of the system).

The second term of the decomposition of (\ref{d7}) over $V_0^{-1}$ defines the
action for local excitations
\be
\label{d14}
p_0d\varphi_0=2V_0\sqrt{\rho_{Un}(\bar\varphi_0)}d\bar\varphi_0+
H_L(\bar\varphi_0)
\frac{d\bar\varphi_0}{\sqrt{\rho_{Un}(\bar\varphi_0)}}
+o\left(\frac{1}{V_0}\right),
\ee
where $\bar\rho_{Un}(\varphi_0)$
is determined by the global equation (\ref{d11}), and
\be
\label{d15}
\frac{d\bar\varphi_0}{\sqrt{\rho_{Un}(\bar\varphi_0)}}
=d\eta_0.
\ee
in accordance with eq.(\ref{d10}).
The reduced action (\ref{d7}) in the zero order in $V_0^{-1}$ in eq. (\ref{d9})
has the form
of conventional field theory  without the global time-reparametrization group
symmetry
\be
\label{d16}
W^R_{(+)}(p_f, f|\bar\varphi_0)=W^G_{(+)}(\bar\varphi_0)+W^L_{(+)}(p_f,
f|\bar\varphi_0),
\ee
where $W^G_{(+)}$ describes evolution of the Universe (see Section 4.) 
and 
\be\label{d17}
W^L_{(+)}(p_f, f |\bar\varphi_0)=\int\limits_{\eta_1}^{\eta_2}d\eta\left(\int
d^3x\sum\limits_fp_fD_{\eta}f-H_L(p_f, f |\bar\varphi(\eta))\right)
\ee
describes local excitations in this Universe.

Really, an observer is using the action for description of
laboratory
experiments in a very small interval of time in the comparison with the
lifetime of the Universe $\eta_0$
\be\label{d18}
\eta_1=\eta_0-\xi\,;\,\eta_2=\eta_0+\xi\,;\,\xi\ll\eta_0,
\ee
and induring this time-interval  $\varphi_0(\eta)$  can be considered
as the constant 
\be\label{d12}
\varphi_0(\eta_0+\xi)\approx\varphi_0(\eta_0)=M_{Pl}\sqrt{\frac{3}{8\pi}}.
\ee
In this case we got the $\sigma-$model version of the standard model
~\cite{PR}.

\section{Conclusion}

In the paper we discussed the status of measurable interval of time ---
``proper time'' in the scheme of the Hamiltonian reduction of GR and
conformal unified theory (CUT) invariant with respect to general coordinate
transformations.

This invariance means that GR and CUT represent an
extended systems (ES) with constraints  and ``superfluous'' variables.
To separate  the physical sector of invariant variables and observables
from parameters of general coordinate transformations, one
needs the procedure of the Hamiltonian reduction which leads to an equivalent
unconstraint system where one of ``superfluous'' variables becomes the
dynamical parameter of evolution.

We have pointed out this ``superfluous''
variable for considered theories (which converts into the evolution parameter
of the reduced system) using the experience of cosmological models and the
Lichnerowich conformally invariant variables.

The dynamics of proper time of an observer  with respect to the
evolution parameter of the reduced system is described by the equation of
ES for the ``superfluous'' canonical momentum.

Just this ``superfluous'' equation of ES determines the ``red
shift'' and Hubble law in cosmological models, GR, and CUT.
To reproduce the Hubble law in quantum theory, the reduced scheme of
quantization of GR and cosmology should be added by the convention of
an observer about measurable time interval.
Normalizability of a wave function is achieved by removing
the ``superfluous'' variable from the set of variables of the reduced system.

From the point of view of the principles of causality and correspondence
with the field theory in the flat space
the considered Hamiltonian reduction of GR prefers to treat
the conformal time as measurable.

We formulated the conformally invariant theory
of fundamental interactions where an observer measures the conformal
time and space intervals. This theory unifies gravitation with the
standard model for strong and electroweak interactions and has
no any dimensional parameters in the Lagrangian. 
In fact, in practice, only the ratios of dimensional quantities are 
the subject of experimental tests. Roughly speaking Planck mass is 
nothing but a multiplicity of the proton mass.

We described the mechanism of appearance of mass scale using as the example 
the dust stage of the evolution of
the Universe and have shown that the value of scalar field at
present time can be determined by the cosmological
data: density of matter and the Hubble constant.
\medskip

{\bf Acknowledgments}

We are happy to acknowledge interesting and critical
discussions with Profs. B.M. Barbashov, R. Brout, A.V. Efremov,
G.A. Gogilidze, V.G. Kadyshevsky, E. Kapu\'scik,
A.M. Khvedelidze, W.Kummer, D. Mladenov,
V.V. Papoyan, and Yu.G. Palii.
We also thank
the Russian Foundation  for Basic
Researches, Grant N 96\--01\--01223
and the Polish
Committee for Scientific Researches, Grant N 603/P03/96, for support.

\end{document}